\documentclass[11pt]{article}
\usepackage[T1]{fontenc}
\usepackage{amsthm}
\usepackage{amsmath}
\usepackage{amsfonts}
\usepackage{graphicx}

\def\be{\begin{equation}}
\def\ee{\end{equation}}
\def\bea{\begin{eqnarray}}
\def\eea{\end{eqnarray}}

\providecommand{\href}[2]{#2}

\newtheorem{defn}{Definition}[section]
\newtheorem{lema}{Lemma}[section]
\newtheorem{coro}{Corollary}[section]
\newtheorem{theo}{Theorem}[section]

\begin{document}
\begin{titlepage}

\title{Linear Hilbertian manifold domains}

\author{Maciej Przanowski\footnote{przan@fis.cinvestav.mx}\\
\it Institute of Physics, Technical University of {\L}\'od\'z,\\
W\'olcza\'nska 219, 93-005 {\L}\'od\'z, Poland\\\\ Marcin
Skulimowski\footnote{mskulim@uni.lodz.pl}\\ \it
Department of Theoretical Physics, University of {\L}\'od\'z,\\
Pomorska 149/153, 90-236 {\L}\'od\'z, Poland
 }


\maketitle \abstract{Necessary and sufficient conditions for a
dense subspace of a Hilbert space to be a linear Hilbertian
manifold domain are given. Some relations between linear
Hilbertian manifold domains and domains of self-adjoint operators
are found.}

\bigskip


\end{titlepage}

\section{Introduction}
Manifold domains play an important role in geometric formulation
of quantum mechanics \cite{cir,sku} and in general infinite
dimensional Hamiltonian systems \cite{che}.
\begin{defn}\label{md}
A subset $D$ of a Banach manifold $M$ is a manifold domain
provided
\begin{itemize}
  \item[1.] $D$ is dense in $M$,
  \item[2.] $D$ carries a Banach manifold structure of its own such
  that the inclusion $i:D\rightarrow M$ is smooth,
  \item[3.] for each $p\in D$, the linear map $T_p i:T_p D\rightarrow T_p
  M$ is a dense inclusion.
\end{itemize}
\end{defn}
\noindent Note that $D$ is not a submanifold of $M$ in the sense
of S.Lang \cite{lan}. In this paper we consider the case when $M$
is a Hilbert space, $D$ is a dense linear subspace of $M$ and $D$
carries a Hilbert space structure of its own. Then it is obvious
that $i:D\rightarrow M$ is linear and if $i:D\rightarrow M$ is
continuous then it is also holomorphic if $M$ is complex and
smooth if $M$ is real. Thus we arrive at the following
\begin{defn}\label{hmd}
A linear subspace $D$ of a Hilbert space $H$ is called a linear
Hilbertian manifold domain of $H$ if
\begin{itemize}
  \item[1.] $D$ is dense in $H$,
  \item[2.] $D$ carries a Hilbert space structure of its own such
  that the inclusion $i:D\rightarrow H$ is continuous.
\end{itemize}
\end{defn}
\noindent (Observe that in the linear case the assumption (3) of
Definition \ref{md} is contained in (1) and (2)).\\ Perhaps, the
most transparent example of a linear Hilbertian manifold domain is
the Sobolev space $W^{m}_2(\Omega)$. Let $\Omega$ be some domain
of $\mathbb{R}^n$ and $L^2(\Omega)$ be the Hilbert space of all
complex functions on $\Omega$ for which $\mid\cdot\mid^2$ is
integrable on $\Omega$. The scalar product and the norm in
$L^2(\Omega)$ are defined by
\begin{equation}
(f,g)_{L^2(\Omega)}:=\int_{\Omega}f(x)\overline{g(x)}dx
\end{equation}
\begin{equation}
\|f\|_{L^2(\Omega)}:=\sqrt{(f,f)_{L^2(\Omega)}},\;\;f,g\in
L^2(\Omega)
\end{equation}
(The overbar stands for the complex conjugation). Sobolev space
$W^m_2(\Omega)$ is defined by
\begin{equation}
W_2^m(\Omega):=\{f\in L^2(\Omega):\;D^{(i_1,...,i_n)}f\in
L^2(\Omega),\;0\leq i_1+...+i_n\leq m\}
\end{equation}
where $D^{(i_1,...,i_n)}f:=\frac{\partial^{i_1+...+i_n}f}{\partial
x_1^{i_1}...\partial x_n^{i_n}}$ and all derivatives are taken in
the sense of the theory of distributions. Then one introduces a
scalar product $(\cdot,\cdot)_{W_2^m(\Omega)}$
\begin{equation}\label{scp}
(u,v)_{W_2^m(\Omega)}:=\sum_{0\leq i_1+...+i_n\leq
m}(D^{(i_1,...,i_n)}u,D^{(i_1,...,i_n)}v)_{L^2(\Omega)}
\end{equation}
for $u,v\in W_2^m(\Omega)$. It can be shown that $W_2^m(\Omega)$
endowed with the scalar product (\ref{scp}) is a separable Hilbert
space. It is also obvious that
\begin{equation}
\|u\|_{L^2(\Omega)}\leq\|u\|_{W_2^m(\Omega)}
\end{equation}
and this leads to the conclusion that the inclusion
$i:W_2^m(\Omega)\rightarrow L^2(\Omega)$ of the Hilbert space
$\left(W^m_2(\Omega),(\cdot,\cdot)_{W^m_2(\Omega)}\right)$ into
$L^2(\Omega)$ is continuous. Finally, since $W^m_2(\Omega)$ is a
dense linear subspace of $L^2(\Omega)$ one concludes that the
Sobolev space $W^m_2(\Omega)$ is a linear Hilbertian manifold domain of
$L^2(\Omega)$. To follow further we consider a Hilbert space
$\widetilde{H}$,
\begin{equation}
\widetilde{H}:=\{(f^{(0,...,0)},...,f^{(i_1,...,i_n)},...,f^{(0,...,m)}):\;0\leq
i_1+...+i_n\leq m,\;f^{(i_1,...,i_n)}\in L^2(\Omega)\}
\end{equation}
\begin{equation}
(F,G)_{\widetilde{H}}:=\sum_{0\leq i_1+...+i_n\leq
m}\left(f^{(i_1,...,i_n)},g^{(i_1,...,i_n)}\right)_{L^2(\Omega)}
\end{equation}
where
$F=(...,f^{(i_1,...,i_n)},...),\;G=(...,g^{(i_1,...,i_n)},...)\in
\widetilde{H}$. Then one defines a linear operator
$D^{(m)}:W^m_2(\Omega)\rightarrow\widetilde{H}$
\begin{equation}
D^{(m)}u:=\left(D^{(0,...,0)}u,...,D^{(i_1,...,i_n)}u,...\right),\;\;0\leq
i_1+...+i_n\leq m,\;\;u\in W^m_2(\Omega).
\end{equation}
It can be shown that the operator $D^{(m)}$ is closed. With the
use of $D^{(m)}$ the scalar product (\ref{scp}) in $W_2^m$ reads
\begin{equation}\label{oper}
(u,v)_{W_2^m(\Omega)}=(D^{(m)}u,D^{(m)}v)_{\widetilde{H}},\;\;u,v\in
W^m_2(\Omega).
\end{equation}
Now the natural question arises if for any linear Hilbertian
manifold domain $D\subset H$ there exists a closed linear operator
$A$ from $D$ to some Hilbert space $\widetilde{H}$ such that the
Hilbert structure on $D$ is defined analogously to (\ref{oper}) by
\begin{equation}\label{oper2}
(\psi,\phi)_D=(A\psi,A\phi)_{\widetilde{H}},\;\;\psi,\phi\in D.
\end{equation}
The answer to this question is the main point of our paper.
\section{Main theorem}
We start with the following
\begin{lema}\label{l1}
Let $H$ be a Hilbert space and $D\subset H$ be a linear dense
domain of $n$ closed linear operators: $A_1:D\rightarrow
H_1,...,A_n:D\rightarrow H_n$ where $H_1,...,H_n$ are Hilbert
spaces. For $1\leq k\leq n$ define a scalar product in $D$ by
\begin{equation}\label{scpk}
(\psi,\phi)^{(k)}_D:=(\psi,\phi)_{H}+\sum_{i=1}^{k}(A_i\psi,A_i
\phi)_{H_i},\;\;\psi,\phi\in D.
\end{equation}
Then $\left(D,(\cdot,\cdot)_D^{(k)}\right)$ is a Hilbert space and
the inclusion $i:D\rightarrow H$ of
$\left(D,(\cdot,\cdot)_D^{(k)}\right)$ into
$\left(H,(\cdot,\cdot)_H\right)$ is continuous, what implies that
$D$ is a linear Hilbertian manifold domain of $H$. Moreover, the
norms $\|\cdot\|_D^{(k_1)}=\sqrt{(\cdot,\cdot)_D^{(k_1)}}$ and
$\|\cdot\|_D^{(k_2)}=\sqrt{(\cdot,\cdot)_D^{(k_2)}}$ are
equivalent for any $1\leq k_1,k_2\leq n$.
\end{lema}
\begin{proof}
Let $\{\phi_j\}_1^{\infty}$, $\phi_j\in D$, be a Cauchy series
with respect to the norm $\|\cdot\|^{(k)}_D$ ($1\leq k\leq n$),
i.e.,
\begin{equation}\label{1}
\lim_{j,l\rightarrow\infty}\|\phi_j-\phi_l\|^{(k)}_D=0.
\end{equation}
From (\ref{scpk}) and the definition of $\|\cdot\|^{(k)}_D$ it
follows that (\ref{1}) yields
\begin{equation}\label{2}
\lim_{j,l\rightarrow\infty}\|\phi_j-\phi_l\|_H=0
\end{equation}
\begin{equation}\label{3}
\lim_{j,l\rightarrow\infty}\|A_i\phi_j-A_i\phi_l\|_{H_i}=0,\;\;i=1,...,k
\end{equation}
what means that
$\{\phi_j\}_1^{\infty},\{A_1\phi_j\}_1^{\infty},...,\{A_k\phi_j\}_1^{\infty}$
are Cauchy series in $H,H_1,...,H_k$ respectively. Hence, there
exist $\phi\in H,\psi_1\in H_1,...,\psi_k\in H_k$ such that
\begin{equation}\label{4}
\lim_{j\rightarrow\infty}\|\phi_j-\phi\|_H=0
\end{equation}
\begin{equation}\label{5}
\lim_{j\rightarrow\infty}\|A_i\phi_j-\psi_i\|_{H_i}=0,\;\;i=1,...,k.
\end{equation}
Since $A_i,...,A_k$ are closed
\begin{equation}
\phi\in D
\end{equation}
and
\begin{equation}
A_i\phi=\psi_i,\;\;i=1,...,k.
\end{equation}
Therefore
\begin{equation}
\left(\|\phi_j-\phi\|^{(k)}_D\right)^2=\|\phi_j-\phi\|^2_H+\sum_{i=1}^k\|A_i(\phi_j-\phi)\|^2_{H_i}\rightarrow
0
\end{equation}
and it means that $\left(D,\|\cdot\|^{(k)}_D\right)$ is complete
for each $1\leq k\leq n$. Consequently
$\left(D,(\cdot,\cdot)^{(k)}_D\right)$ is a Hilbert space for any
$1\leq k\leq n$. Obviously
\begin{equation}\label{6}
\|\psi\|_H\leq\|\psi\|^{(k)}_D,\;\;\psi\in D,\;\;k=1,...,n.
\end{equation}
From (\ref{6}) one gets that the inclusion $i:D\rightarrow H$ is a
continuous mapping from $\left(D,(\cdot,\cdot)^{(k)}_D\right)$
into $\left(H,(\cdot,\cdot)_H\right)$. Thus $D$ is a linear
Hilbertian manifold domain of $H$. Moreover
\begin{equation}
\|\cdot\|^{(k_1)}_D\leq\|\cdot\|^{(k_2)}_D
\end{equation}
for any $k_1\leq k_2$. Therefore, employing {\em the open mapping
theorem} \cite{yos} one concludes that the norms
$\|\psi\|^{(k_1)}_D$ and $\|\psi\|^{(k_2)}_D$ are equivalent. The
proof is complete.
\end{proof}
\noindent Observe that defining
\[\widetilde{H}^{(k)}=H\oplus H_1\oplus...\oplus H_k\]
\begin{equation}\label{7}
\widetilde{A}^{(k)}:D\rightarrow\widetilde{H}^{(k)},\;\;\widetilde{A}^{(k)}(\psi):=\psi\oplus
A_1\psi\oplus...\oplus A_k\psi,\;\;\psi\in D,\;1\leq k\leq n.
\end{equation}
we can rewrite (\ref{scpk}) in the following compact form
\begin{equation}\label{8}
\left(\psi,\phi\right)_D^{(k)}=\left(\widetilde{A}^{(k)}\psi,\widetilde{A}^{(k)}\phi\right)_{\widetilde{H}^{(k)}}.
\end{equation}
Of course $\widetilde{A}^{(k)}:D\rightarrow\widetilde{H}^{(k)}$ is
a closed linear operator. From Lemma \ref{l1} one easily obtains
\begin{coro}{\normalfont\cite{cir}}\label{c1}
If $D_A\subset H$ is a dense domain of a self-adjoint operator
$A:D_A\rightarrow H$, $A=A^{*}$, then
$\left(D_A,(\cdot,\cdot)_{D_A}\right)$ with
\begin{equation}\label{9}
(\phi,\psi)_{D_A}:=(\phi,\psi)_{H}+(A\phi,A\psi)_{H}
\end{equation}
is a Hilbert space and the inclusion $i:D\rightarrow H$ is
continuous, what implies that $D_A$ is a linear Hilbertian
manifold domain of $H$.
\end{coro}
\begin{proof}
$A=A^{*}\;\Longrightarrow\;A$ is closed. Then from Lemma~\ref{l1}
we get the thesis.
\end{proof}
\noindent The fact that any dense domain of a self-adjoint
operator is a linear Hilbertian manifold domain appears to be
fundamental in geometrical formulation of quantum mechanics (see
\cite{cir}). Before giving the main theorem we prove a simple
lemma
\begin{lema}\label{l22}
Let $\left(H,(\cdot,\cdot)_H\right)$ be a Hilbert space and
$D\subset H$ be a dense linear subspace of $H$. If there exists a
scalar product $(\cdot,\cdot)_D$ in $D$ such that
$\left(D,(\cdot,\cdot)_D\right)$ is a separable Hilbert space and
$\|f\|_H\leq a\|f\|_D$ $\forall f\in D$, $a>0$, then
\begin{equation}\label{10}
dim\left(H,(\cdot,\cdot)_H\right)=dim\left(D,(\cdot,\cdot)_D\right).
\end{equation}
\end{lema}
\begin{proof}
For $dim\left(D,(\cdot,\cdot)_D\right)<\infty$ the proof is
trivial. Let $dim\left(D,(\cdot,\cdot)_D\right)=\aleph_0$, where
$\aleph_0$ is the cardinal number of the set of natural numbers
$\mathbb{N}$. From the assumption that the norm $\|\cdot\|_D$ is
stronger than $\|\cdot\|_H$ and $D$ is dense in
$\left(H,(\cdot,\cdot)_H\right)$ it follows that every set which
is dense in $\left(D,(\cdot,\cdot)_D\right)$ is also dense in
$\left(H,(\cdot,\cdot)_H\right)$. Since
$\left(D,(\cdot,\cdot)_D\right)$ is separable there exists a
countable dense set $S$ in $\left(D,(\cdot,\cdot)_D\right)$. Then
$S$ is also dense in $\left(H,(\cdot,\cdot)_H\right)$. Thus
$\left(H,(\cdot,\cdot)_H\right)$ is separable and (\ref{10}) holds
true.
\end{proof}
\noindent From Lemma~\ref{l22} one gets
\begin{coro}\label{c2}
If a linear Hilbertian manifold domain $D$ of $H$ admits a scalar
product $(\cdot,\cdot)_D$ such that
$\left(D,(\cdot,\cdot)_{D}\right)$ is a separable Hilbert space
and $\|f\|_H\leq\|f\|_D$ then the Hilbert space
$\left(H,(\cdot,\cdot)_{H}\right)$ is also separable.\qed
\end{coro}
\noindent Now we are in a good position to prove the main theorem:
\begin{theo}\label{main}
Let $\left(H,(\cdot,\cdot)_H\right)$ be a Hilbert space and let
$D\subset H$ be a dense linear subspace of $H$. If there exists a
linear operator $A:D\rightarrow H$ satisfying the following
conditions
\begin{itemize}
  \item[(a)] A is one-to-one and $A(D)=H$,
  \item[(b)] The inverse operator $A^{-1}:H\rightarrow D$ is
  bounded,
\end{itemize}
then $D$ is a linear Hilbertian manifold domain of $H$ and
\begin{equation}\label{11}
(f,g)_D:=\left(A\circ i f,A\circ i g\right)_H,\;\;f,g\in D,
\end{equation}
where $i:D\rightarrow H$ is the inclusion of $D$ into $H$, defines
a scalar product in $D$ such that $\left(D,(\cdot,\cdot)_D\right)$
is a Hilbert space and the inclusion $i:D\rightarrow H$ of the
Hilbert space $\left(D,(\cdot,\cdot)_D\right)$ into
$\left(H,(\cdot,\cdot)_H\right)$ is continuous.\\ Conversely, if
$D\subset H$ is a linear Hilbertian manifold domain of $H$ and
$(\cdot,\cdot)_D$ is any scalar product in $D$ such that
$\left(D,(\cdot,\cdot)_D\right)$ is a separable Hilbert space and
the inclusion $i:D\rightarrow H$ is continuous then there exists a
linear operator $A:D\rightarrow H$ satisfying (a) and (b) and such
that the scalar product $(\cdot,\cdot)_D$ is given by (\ref{11}).
Any linear operator satisfying (a) and (b) is closed.
\end{theo}
\begin{proof}
Let $A:D\rightarrow H$ be a linear operator fulfilling (a) and
(b). One easily shows that the formula (\ref{11}) defines a scalar
product $(\cdot,\cdot)_D$ in $D$. We prove that
$\left(D,(\cdot,\cdot)_D\right)$ is a Hilbert space. Let
$\{f_n\}_1^{\infty}$, $f_n\in D$, be a Cauchy series in
$\left(D,(\cdot,\cdot)_D\right)$ i.e.,
\begin{equation}\label{12}
\lim_{n,m\rightarrow\infty} \| f_n-f_m {\|}_D=0.
\end{equation}
Denote $H\ni h_n:=A\circ i f_n$. By (\ref{11}) and (\ref{12})
\[\lim_{n,m\rightarrow\infty} \| h_n-h_m {\|}_H=0.\]
This means that $\{h_n\}_1^{\infty}$ is a Cauchy series in $H$
and, consequently, there exists $h\in H$ such that
\[\lim_{n,m\rightarrow\infty} \| h_n-h {\|}_H=0.\]
Define $f:=i^{-1}\circ A^{-1}h\in D$. Then
\[\| f_n-f{\|}_D=\| h_n-h{\|}_H\rightarrow 0.\]
Hence $f\in D$ is the limit of the Cauchy series
$\{f_n\}_1^{\infty}$. This proves the completeness of
$\left(D,(\cdot,\cdot)_D\right)$ and one concludes that
$\left(D,(\cdot,\cdot)_D\right)$ is a Hilbert space. For any
vector $g$ of this Hilbert space
\begin{equation}\label{13}
\|g\|_D=\|A\circ i g\|_H.
\end{equation}
By the condition (b)
\begin{equation}\label{14}
\|i g\|_H=\|A^{-1}\circ A\circ i g\|_H\leq\|A^{-1}\|\|A\circ i
g\|_H\Rightarrow\|i g\|_H\leq\|A^{-1}\|\|g\|_D.
\end{equation}
Therefore, the norm $\|\cdot\|_D$ is stronger than $\|\cdot\|_H$
and the inclusion $i:D\rightarrow H$ is continuous. This ends the
proof of the first part of our theorem.\\ To prove the second
part, first note that under the assumption that the Hilbert space
$\left(D,(\cdot,\cdot)_D\right)$ is separable Lemma~\ref{l22}
gives
\[dim\left(D,(\cdot,\cdot)_D\right)=dim\left(H,(\cdot,\cdot)_H\right)=\aleph_0.\]
Hence there exists an isometry $U:H\rightarrow D$ of
$\left(H,(\cdot,\cdot)_H\right)$ onto
$\left(D,(\cdot,\cdot)_D\right)$ \cite{yos,gla}
\begin{equation}\label{15}
\left(U h_1,U h_2\right)_D=\left(h_1,h_2\right)_H,\;\;h_1,h_2\in H.
\end{equation}
Then the linear operator $i\circ U:H\rightarrow D\subset H$ is a
one-to-one bounded (with respect to $\|\cdot\|_H$) linear operator
from $H$ onto $D$. Consequently, the inverse operator
\[\left(i\circ U\right)^{-1}=:A:D\rightarrow H\]
is a one-to-one linear operator from $D$ onto $H$ and obviously it
satisfies the conditions (a) and (b). Then it is easy to see that
(\ref{15}) leads to (\ref{11}). This proves the second part of the
theorem.\\ Finally, since $A^{-1}:H\rightarrow D$ is by (b) a
bounded linear operator, it is also closed. Hence, $A:D\rightarrow
H$ being an inverse operator to the closed linear operator
$A^{-1}$ is also closed \cite{gla}. The proof is complete.
\end{proof}
\noindent The formula (\ref{11}) corresponds to the formulas
(\ref{oper}), (\ref{oper2}) or (\ref{8}) but now the Hilbert
spaces $\widetilde{H}$ or $\widetilde{H}^{(k)}$ are exactly the
original Hilbert space $H$ to which the linear Hilbertian manifold
domain $D$ belongs. From Theorem~\ref{main} one concludes that:\\\\
{\em Every linear Hilbertian manifold domain $D$ of $H$, admitting
a scalar product $(\cdot,\cdot)_D:D\times D\rightarrow \mathbb{C}$
such that $\left(D,(\cdot,\cdot)_D\right)$ is a separable Hilbert
space and the inclusion $i:D\rightarrow H$ is continuous, is the
domain of some linear operator $A:D\rightarrow H$ satisfying the
conditions
(a) and (b). These conditions imply that $A$ is closed.}\\\\
\noindent Corollary~\ref{c1} tells us that any dense domain of a
self-adjoint operator is a linear Hilbertian manifold domain. We
are not able to prove if the inverse statement is also true.
Nevertheless, in the next section we prove some weakenned form of
this statement.
\section{Self-adjoint operators and linear Hilbertian manifold domains}
Let $H$ be a separable Hilbert space and $D\subset H$ be a linear
Hilbertian manifold domain of $H$. Employing Theorem~\ref{main}
one can easily show that there exists a linear Hilbertian manifold
domain $D_A$ of $H$, $D_A\subset D\subset H$, and an operator
$A:D_A\rightarrow H$ satisfying $(a)$ and $(b)$. Since $A$ is a
closed linear operator and $D_A$ is dense in $H$ the operator
$B:=A^{*}A$ is self-adjoint and positive, and the domain of $B$,
$D_B\subset D_A\subset D\subset H$ is dense in $H$ (see
\cite{gla}). $A^{-1}:H\rightarrow D_A$ is a one-to-one and bounded
linear operator from $H$ onto $D_A$ which implies that
$(A^{-1})^{*}:H\rightarrow D^{*}_{A}$ is also a one-to-one,
bounded linear operator from $H$ onto $D^{*}_{A}$. It can be
proved \cite{gla} that $(A^{-1})^{*}=(A^{*})^{-1}$. Consequently,
the linear operator $B=A^{*}A:D_B\rightarrow H$ is a one-to-one,
self adjoint and positive operator from $D_B$ onto $B(D_B)$. We
define scalar product $(\cdot,\cdot)_B$ in $D_B$ by
\begin{equation}\label{17}
(f,g)_{D_B}:=(Bf,Bg)_H,\;\;f,g\in D_B
\end{equation}
One can show that
\begin{equation}\label{18}
\|f\|_{D_B}\geq\frac{1}{\|A^{-1}\|^2}\|f\|_H,\;\;f\in D_B
\end{equation}
where as usually $\|f\|_{D_B}=\sqrt{(f,f)_{D_B}}$.\\ Let
$\{f_n\}_1^{\infty}$, $f_n\in D_B$, be a Cauchy series in
$\left(D_B,(\cdot,\cdot)_{D_B}\right)$ i.e.,
\begin{equation}\label{19}
\lim_{n,m\rightarrow\infty} \| f_n-f_m {\|}_{D_B}=0
\end{equation}
By (\ref{18}) and (\ref{19}) we get
\begin{equation}\label{20}
\lim_{n,m\rightarrow\infty} \| f_n-f_m {\|}_{H}=0
\end{equation}
Hence, there exists $f\in H$ such that
\begin{equation}\label{21}
\lim_{n\rightarrow\infty} \| f_n-f{\|}_{H}=0
\end{equation}
But from (\ref{17}) and (\ref{19}) it follows that there exists
$g\in H$ such that
\begin{equation}\label{22}
\lim_{n\rightarrow\infty} \|B f_n-g{\|}_{H}=0
\end{equation}
Since $B:D_B\rightarrow H$ is self-adjoint it is also closed and,
consequently, $f\in D_B$ and $g=Bf$. Finally, one concludes that
\begin{equation}\label{23}
\lim_{n\rightarrow\infty} \|
f_n-g{\|}_{D_B}=\lim_{n\rightarrow\infty} \|B f_n-B f{\|}_{H}=0
\end{equation}
and it means that $\left(D_B,(\cdot,\cdot)_{D_B}\right)$ is a
Hilbert space. Moreover, by (\ref{18}) the inclusion
$i:D_B\rightarrow H$ is continuous. Thus we arrive at the
following
\begin{theo}\label{fin}
Let $D$ be a linear Hilbertian manifold domain of a separable
Hilbert space $\left(H,(\cdot,\cdot)_{H}\right)$. Then there
exists a dense linear subspace $D_B$ of $H$, $D_B\subset D\subset
H$ and a self-adjoint positive linear operator $B:D_B\rightarrow
H$ such that $\left(D_B,(\cdot,\cdot)_{D_B}\right)$ is a Hilbert
space with
\[(f,g)_{D_B}:=(Bf,Bg)_H,\;\;f,g\in{D_B}\]
and the inclusion $i:D_B\rightarrow H$ is a continuous inclusion
of $\left(D_B,(\cdot,\cdot)_{D_B}\right)$ into
$\left(H,(\cdot,\cdot)_{H}\right)$.\qed
\end{theo}
\noindent Analogously as in Lemma 2.1 one gets that the scalar
product $<\cdot,\cdot>_{D_B}$ in $D_B$ defined by
\begin{equation}\label{24}
<f,g>_{D_B}=(f,g)_H+(Bf,Bg)_H,\;\;f,g\in D_B
\end{equation}
leads to the norm in $D_B$ equivalent to that given by the scalar
product $(\cdot,\cdot)_{D_B}$.\\ Concluding one can say that:\\
{\em If $D$ is a linear Hilbertian manifold domain of a separable
Hilbert space $H$ then there exists a linear Hilbertian manifold
domain $D_B\subset D\subset H$ of $H$ which is the domain of some
self-adjoint positive linear operator $B:D_B\rightarrow H$.}\\
\noindent This conclusion is a weakened form of the statement
inverse to Corollary~\ref{c1}.
\section{Manifold domains and observables in geometric quantum mechanics}
In ordinary quantum mechanics observables are represented by
densely defined self-adjoint linear operators in a separable
Hilbert space $H$. As is well known if
$\mathcal{O}:D_{\mathcal{O}}\rightarrow H$ is a self-adjoint
linear operator from $D_{\mathcal{O}}\subset H$ to $H$ then
\begin{equation}\label{41}
\exp\{it\mathcal{O}\}:=\int^{+\infty}_{-\infty}\exp\{it\lambda\}dE_{\lambda},\;\;t\in(-\infty,+\infty)
\end{equation}
(where $E_{\lambda}$, $\lambda\in (-\infty,+\infty)$ is the
spectral measure associated with $\mathcal{O}$) defines a strongly
continuous one-parameter group of unitary operators on $H$.
Conversely, by Stone's theorem, every strongly continuous
one-parameter group of unitary operators on $H$ is given by
(\ref{41}). Hence, in quantum mechanics observables be eventually
indentified  with strongly continuous one-parameter groups of
unitary operators on $H$. In geometric quantum mechanics
\cite{cir,sku,cii,ana,ash,bro} an observable is represented by the
so called {\em observable function}
\begin{equation}\label{42}
H\supset D_{\mathcal{O}}\ni f\rightarrow \langle
\mathcal{O}\rangle (f):=\left(\mathcal{O}f,f\right)_H\in R.
\end{equation}
From Corollary~\ref{c1} it follows that $D_\mathcal{O}$ is a
linear Hilbertian manifold domain of $H$. Then the Hamiltonian
vector field $X$ on $D_{\mathcal{O}}$ defined by the observable
function (\ref{42})
\begin{equation}\label{43}
X_f\lrcorner\;\omega=d\langle\mathcal{O}\rangle (f),\;\;f\in
D_{\mathcal{O}}
\end{equation}
where $\omega$ is the natural symplectic form on $H$
\begin{equation}
\omega(f,g):=-2Im(f,g),\;\;f,g\in H
\end{equation}
generates (uniquely) the one-parameter continuous group
$\mathbb{R}\times H\ni (t,g)\longmapsto \varphi_t(g)\in H$ of
K\"{a}hler isomorphisms of $H$ satisfying
\begin{equation}\label{45}
\frac{d\varphi_t(f)}{dt}|_{t=0}=X_f,\;\;f\in D_{\mathcal{O}}.
\end{equation}
(Recall that a K\"{a}hler isomorphism of $H$ is such an
isomorphism which preserves the K\"{a}hler structure on $H$)\\
\noindent Conversely, again by Stone's theorem, every
one-parameter continuous group $\mathbb{R}\times H\ni
(t,g)\longmapsto \varphi_t(g)\in H$ of K\"{a}hler isomorphisms of
$H$ defines (uniquely) an observable function
$\langle\mathcal{O}\rangle$ determined on some linear Hilbertian
manifold domain $D_{\mathcal{O}}\subset H$ according to
(\ref{42}). The Hamiltonian vector field $X$
on $D_{\mathcal{O}}$ given by (\ref{43}) satisfies the relation (\ref{45}).\\
\noindent\underline{Remark:} All that can be easily carry over to
the projective space $PH$ which is the true phase space in quantum
theory (see \cite{cir,sku} and \cite{cii,ana,ash,bro} for
details). We don't consider this problem as we deal with linear
domains of Hilbert
space $H$.\\\\
From Theorem~\ref{fin} one can extract an interesting
conclusion:\\
{\em For any linear Hilbertian manifold domain $D$ of a separable
Hilbert space $H$ there exists a positive observable function
$\langle B\rangle:D_B\rightarrow \mathbb{R}$ ($\langle
B\rangle(f)\geq 0\;\forall f\in D_B$; $\langle B\rangle(f)=0
\Longleftrightarrow f=0$) where $D_B\subset D\subset H$ is some
linear Hilbertian manifold domain.}\\\\
{\bf Acknowledgmens}\\\\
We are indebted to Bogdan Nowak for valuable discussion. One of us
(Maciej Przanowski) was partly supported by the NATO grant
PST.CLG.978984.

\end{document}